\documentclass[pre,preprint,showpacs]{revtex4}%
\usepackage{amsfonts}
\usepackage{amsmath}
\usepackage{amssymb}
\usepackage{graphicx}%
\begin{document}
\title{Fluctuation Theorems for Systems under Fokker-Planck dynamics}
\author{A. P\'{e}rez-Madrid}
\affiliation{Departament de F\'{\i}sica Fonamental, Facultat de F\'{\i}sica, Universitat de
Barcelona, Av. Diagonal 647, 08028 Barcelona, Spain}
\author{I. Santamar\'{\i}a-Holek}
\affiliation{Facultad de Ciencias, Universidad Nacional Aut\'{o}noma de M\'{e}xico,
Circuito exterior de Ciudad Universitaria 04510, D.F., M\'{e}xico}

\begin{abstract}
We study Brownian motion driven with both conservative and nonconservative
external forces. By using the thermodynamic approach of the theory of Brownian
motion we obtain the Fokker-Planck equation and derive expressions for the
Fluctuation Theorem in local equilibrium and in quasi-equilibrium. In local
equilibrium the expressions we obtain coincide with previous results.

\end{abstract}
\startpage{1}
\endpage{102}
\maketitle

\section{Introduction}

Since the first formulation of a Fluctuation Theorem (FT) \cite{evans2} there
has been an increasing interest in this subject. Further contributions
\cite{cohen} have broadened the scope of the physical situations where a FT
arises. An extensive account of the literature can be found in Refs.
\cite{evans} and \cite{searles}. The search for certain symmetries or rules in
the different systems or situations we can find far from equilibrium is the
major element of interest in the different expressions of the FT. It seems
that the FT play a role similar to the partition function in equilibrium
statistical mechanics.

These theorems are concerned with the relation between nonequilibrium
measurements and certain thermodynamic magnitudes such as entropy, heat and
free energies, \cite{searles}\textbf{.} They involve pairs of nonequilibrium
trajectories in the phase space, one and its reverse, and their corresponding probabilities.

Our contention here is to clarify the thermodynamical roots of the FT,
determining their scope for systems with Fokker-Planck dynamics. We will
obtain the expression of the FT in local equilibrium \ and in states of quasi-equilibrium.

The paper is organized as follows. In section 2, we perform the thermodynamic
analysis of an ensemble of non-interacting Brownian particles under the action
of both conservative and nonconservative forces, deriving the Fokker-Planck
equation. In section 3, we derive the configurational relaxation equations
when the system is in local equilibrium and in quasi-equilibrium and obtain
the parameters which characterize these states. Section 4 is devoted to the
derivation of the FTs. Finally in section 5 we formulate our main conclusions.

\section{Thermodynamic analysis}

Consider an ensemble of non-interacting Brownian particles subjected to a
potential field $V(x)$ which is initially in equilibrium with a heat bath at
temperature $T_{o}$. In the canonical ensemble, the system is distributed
according to\textbf{ }
\begin{equation}
\rho_{eq.}\sim\exp\left[  -\frac{H\left(  \Gamma\right)  }{k_{B}T_{o}}\right]
,\label{density}%
\end{equation}
where $H\left(  \Gamma\right)  =(1/2)mu^{2}+mV(x)$ is the Hamiltonian and
$\Gamma=(x,u)$ represents a point of the one-particle phase space spanned by
the position $x$ and the velocity $u$ of the particle.

Let us assume that at certain moment ($t=0$), a nonconservative force $f(t)$
is applied on the system. Then, the system evolves in time and can be
described by using the nonequilibrium probability density $\rho(\Gamma,t)$ and
the nonequilibrium entropy of the system which is given through the Gibbs
entropy postulate \cite{kampen},\cite{mazur}%
\begin{equation}
S(t)=-k_{B}\int\rho(\Gamma,t)\ln\frac{\rho(\Gamma,t)}{\rho_{eq.}}%
d\Gamma+S_{eq.}. \label{non-eq-entropy}%
\end{equation}
where $S_{eq.}$ is the equilibrium entropy of the Brownian particles plus the
bath. The variations in the probability density $\rho(\Gamma,t)$, imply
changes in the nonequilibrium entropy which can be obtained from Eq.
(\ref{non-eq-entropy})
\begin{equation}
\delta S=-\frac{1}{T_{o}}\int\mu(\Gamma,t)\delta\rho(\Gamma,t)d\Gamma.
\label{gibbs-eq}%
\end{equation}
The thermodynamically conjugated nonequilibrium chemical potential $\mu
(\Gamma,t)$ of the density $\rho(\Gamma,t)$ has been defined by
\begin{equation}
\mu(\Gamma,t)=k_{B}T_{o}\ln\frac{\rho(\Gamma,t)}{\rho_{eq.}}+\mu_{eq.},
\label{noeq-chem-pot}%
\end{equation}
where $\mu_{eq}$ is the equilibrium chemical potential. Equation
(\ref{gibbs-eq}) is similar to the Gibbs' equation of thermodynamics
\cite{vilar}, \cite{reguera}.

After the force $f(t)$ has been applied, the variation in time of $\rho
(\Gamma,t)$ is given by the generalized Liouville equation%
\begin{gather}
\frac{\partial}{\partial t}\rho(\Gamma,t)+V_{\Gamma}(\Gamma,t)\cdot
\nabla_{\Gamma}\rho(\Gamma,t)\nonumber\\
=-\frac{\partial}{\partial u}\rho(\Gamma,t)f(t)-\frac{\partial}{\partial
u}J(\Gamma,t), \label{gen-liouville}%
\end{gather}
where $V_{\Gamma}(\Gamma,t)=\left(  u,-\nabla V(x)\right)  $ is the phase
space velocity corresponding to the Hamiltonian flow,\textbf{ $\nabla_{\Gamma
}=(\nabla,\partial/\partial u)$ }with $\nabla$ the spatial derivative and
$J(\Gamma,t)$ constitutes a diffusion current in phase space. By using Eqs.
(\ref{gibbs-eq}) and (\ref{gen-liouville}) and performing partial integrations
assuming that the currents vanish at boundaries, one obtains that the total
rate of change of the nonequilibrium entropy (\ref{non-eq-entropy}) is
\begin{equation}
\frac{dS}{dt}=-\frac{1}{T_{o}}\left\langle f(t)u\right\rangle -\frac{1}{T_{o}%
}\int J(\Gamma,t)\frac{\partial}{\partial u}\mu(\Gamma,t)d\Gamma,
\label{ent-chan}%
\end{equation}
where the quantity $\langle f(t)u\rangle=\int f(t)u\rho(\Gamma,t)d\Gamma$
constitutes the power, $dw/dt$, supplied by the external force $f(t)$ which is
dissipated as heat into the bath. Thus, assuming that \cite{zwanzig}
\begin{equation}
\rho(\Gamma,t)=\delta\left(  x(t)-x\right)  \delta\left(  \dot{x}(t)-u\right)
, \label{path density}%
\end{equation}
we can write%
\begin{equation}
\frac{dw}{dt}=f(t)\dot{x}(t), \label{heat}%
\end{equation}
where $dw$ is the amount of work done on the system in a time $dt$ by the
external force $f$.

The second term on the right-hand side of Eq. (\ref{ent-chan}), represents the
change of the entropy due to a gradient of the chemical potential and
therefore constitutes the entropy production $\sigma$ due to diffusion in the
phase space
\begin{equation}
\sigma=-\frac{1}{T_{o}}\int J(\Gamma,t)\frac{\partial}{\partial u}\mu
(\Gamma,t)d\Gamma. \label{ent-pro}%
\end{equation}
This quantity accounts for the internal dissipative processes and it is
assumed, according to the second law of thermodynamics, that it does not take
negative values: $\sigma\geq0$, \cite{mazur}.

Hence, by introducing the exchange of entropy due to the interactions with the
environment as\textbf{ }
\begin{equation}
\frac{d_{ext}S}{dt}=\frac{1}{T_{o}}\frac{dw}{dt}, \label{ext-entr}%
\end{equation}
which has an undefined sign, we finally obtain
\begin{equation}
\frac{dS}{dt}=\frac{d_{ext}S}{dt}+\sigma\label{rate-entropy}%
\end{equation}
which expresses the entropy balance between the exchange of entropy with the
bath and the entropy generated in the irreversible processes established in
the system when removed from its equilibrium state.

To completely characterize the stochastic dynamics of the system, we must find
the expression of the current $J(\Gamma,t)$. According to Eq. (\ref{ent-pro})
and following the rules of nonequilibrium thermodynamics \cite{mazur}, this
can be achieved by establishing a linear relation between the current
$J(\Gamma,t)$ and their conjugated thermodynamic force $\partial\mu/\partial
u$. This relation can be expressed by
\begin{equation}
J(\Gamma,t)=-\frac{L}{T_{o}}\frac{\partial}{\partial u}\mu(\Gamma,t),
\label{phenomeno}%
\end{equation}
where $L$ is a phenomenological coefficient. By using the expression of the
chemical potential given through Eq. (\ref{noeq-chem-pot}) in (\ref{phenomeno}%
), one obtains%
\begin{equation}
J(\Gamma,t)=-\gamma\left(  \frac{k_{B}T_{o}}{m}\frac{\partial}{\partial
u}+u\right)  \rho(\Gamma,t), \label{current}%
\end{equation}
where we have identified $mL/\rho T_{o}=\gamma$, with $\gamma$ being the
friction coefficient per unit mass of the Brownian particles. By substituting
now Eq.(\ref{current}) into Eq. (\ref{gen-liouville}) we obtain
\begin{align}
\frac{\partial}{\partial t}\rho &  =-\nabla u\rho+\frac{\partial}{\partial
u}\left[  \nabla V(x)-f(t)\right]  \rho\nonumber\\
&  +\gamma\frac{\partial}{\partial u}\left(  \frac{k_{B}T_{o}}{m}%
\frac{\partial}{\partial u}+u\right)  \rho\label{fokker-planck}%
\end{align}
which is the Fokker-Planck (or Klein-Kramers) equation governing the time
evolution of the nonequilibrium probability density $\rho(\Gamma,t)$ in the
presence of a conservative potential $V(x)$ and a non-conservative force
$f(t)$\textbf{.}

\section{Local equilibrium and quasiequilibrium}

When the dynamics of the system can be characterized by two or more time
scales, its long-time behavior can manifest different dynamical regimes
depending on the existence (or not) of a local equilibrium state. Here, we
will show the conditions that must be satisfied by the system in order to
reach a local equilibrium state and how, when these conditions are not
satisfied, the system enters in a quasiequilibrium state leading to a
different long-time dynamical behavior\textbf{.}

Assuming the presence of two time scales in which $u$ is the fast variable,
the dynamical processes associated to configurational changes in the system
are related to the slow variable $x$. Thus, the long-time behavior of the
system can be more conveniently described by writing \cite{agusti}\textbf{ }
\begin{equation}
\rho(\Gamma,t)=\phi_{x}(u,t)n(x,t), \label{decoupling}%
\end{equation}
where $\phi_{x}(u,t)$, is a conditional probability density and $n(x,t)=\int
\rho(\Gamma,t)du$ is a reduced probability density in $x$-space which evolves
according to%
\begin{equation}
m\frac{\partial}{\partial t}n(x,t)=-\nabla\int mu\rho(\Gamma,t)du.
\label{continuity}%
\end{equation}
This equation can be obtained by integration over $u$ of Eq.
(\ref{fokker-planck}) and thus implicitly defines the current $J(x,t)=\int
mu\rho(\Gamma,t)du$.

i) \textit{Local equilibrium}.- In order to achieve a local equilibrium state,
the system must satisfy that $\phi_{x}(u,t)\sim\exp\left(  -mu^{2}/2k_{B}%
T_{o}\right)  $, i.e. that the distribution of velocities is given by the
equilibrium Maxwellian. In this case, by multiplying Eq. (\ref{fokker-planck})
by $mu$ and performing a partial integration over $u$-space, we obtain the
evolution equation for $J(x,t)$%
\begin{gather}
\frac{\partial}{\partial t}J(x,t)+\gamma J(x,t)=-mn(x,t)\left[  \nabla
V(x)-f(t)\right]  \nonumber\\
-k_{B}T_{o}\nabla n(x,t).\label{cur-balance}%
\end{gather}
For times $t\gg\gamma^{-1}$ Eq. (\ref{cur-balance}) gives
\begin{equation}
J(x,t)\simeq-\gamma^{-1}\left\{  mn(x,t)\left[  \nabla V(x)-f(t)\right]
+k_{B}T_{o}\nabla n(x,t)\right\}  .\label{current-2}%
\end{equation}
In order that our results be valid, here and henceforth we will assume that
the time scale over which $f(t)$ varies should be larger than $\gamma^{-1}$.
Otherwise, the time derivative of Eq. (\ref{cur-balance}) must be taken into
account. Once we substitute Eq. (\ref{current-2}) into Eq. (\ref{continuity})
we obtain the Smoluchowski equation in the presence of external forces%
\begin{equation}
\frac{\partial}{\partial t}n(x,t)=D\nabla\left\{  \frac{m}{k_{B}T_{o}%
}n(x,t)\left[  \nabla V(x)-f(t)\right]  +\nabla n(x,t)\right\}  \label{Smolu}%
\end{equation}
with $D=k_{B}T_{o}/m\gamma$ being the diffusion coefficient. The Smoluchowski
equation (\ref{Smolu}) admits a local equilibrium solution given by
\begin{equation}
n_{l.eq.}(x;f)\sim\exp\left[  -\frac{m}{k_{B}T_{o}}\int^{x}\left[  \nabla
V(x^{\prime})-f(t)\right]  dx^{\prime}\right]  .\label{eq-sol}%
\end{equation}
Hence, the nonequilibrium entropy given through Eq. (\ref{non-eq-entropy})
reduces to%
\begin{equation}
S(t)=-k_{B}\int n(x,t)\ln\frac{n(x,t)}{n_{l.eq.}}dx+S_{l.eq.}%
(t),\label{non-eq-entropy2}%
\end{equation}
where $S_{l.eq.}(t)=-k_{B}\int n(x,t)\ln[n_{l.eq.}(x;f)/{n_{eq.}}]dx+S_{eq.}$
is the local equilibrium entropy. It is important to emphasize that through
Eq. (\ref{eq-sol}), the local equilibrium state is characterized by a
probability density containing the temperature of the bath $T_{o}$.

ii) \textit{Quasi-equilibrium}.- Released from the restrictive condition of
local equilibrium, instead of obtaining the expression (\ref{current-2}) for
the diffusion current $J(x,t)$, at long time we find%
\begin{equation}
J(x,t)=-\gamma^{-1}\left\{  mn(x,t)\left[  \nabla V(x)-f(t)\right]
+k_{B}\nabla n(x,t)T(x,t)\right\}  ,\label{current3}%
\end{equation}
where the local temperature of the system $T(x,t)$ has been defined through a
generalization of the equipartition theorem as the second moment of $\phi
_{x}(u,t)$ \cite{landau1},\cite{pathria}
\begin{equation}
k_{B}T(x,t)=m\int u^{2}\phi_{x}(u,t)du.\label{temperature}%
\end{equation}
The current given in Eq. (\ref{current3}) can be rewritten as
\begin{equation}
J(x,t)=-D(x,t)\left[  \nabla n(x,t)+\frac{mn(x,t)}{k_{B}T(x,t)}\left[
\nabla\Phi(x,t)-f(t)\right]  \right]  ,\label{current4}%
\end{equation}
where $\Phi(x,t)=V(x)+k_{B}T(x,t)$ is an effective potential and
$D(x,t)=k_{B}T(x,t)m^{-1}\gamma^{-1}$ is the bare effective diffusion
coefficient. Hence, by substituting Eq. (\ref{current4}) into Eq.
(\ref{continuity}) we obtain%
\begin{equation}
\frac{\partial}{\partial t}n(x,t)=\nabla\left\{  D(x,t)\left[  \nabla
n(x,t)+\frac{mn(x,t)}{k_{B}T(x,t)}\left[  \nabla\Phi(x,t)-f(t)\right]
\right]  \right\}  .\label{Smolu2}%
\end{equation}
The diffusion equation (\ref{Smolu2}) given above admits a quasi-equilibrium
solution
\begin{equation}
n_{qe.}(x\mathbf{,}f)\sim\exp\left[  -\int^{x}\frac{m}{k_{B}T(x^{\prime}%
,f)}\left[  \nabla\Phi(x^{\prime},f)-f(t)\right]  dx^{\prime}\right]
\label{quasi-equi}%
\end{equation}
obtained when the probability current instantaneously vanishes.

An estimation of the temperature $T$ independent of the position can be
obtained by first deriving the evolution equation of the temperature field
$T(x,t)$. Thus, by multiplying Eq. (\ref{fokker-planck}) by $mu^{2}$ and
integrating in $u$ one has
\begin{gather}
\frac{1}{2}k_{B}\frac{\partial}{\partial t}n(x,t)T(x,t)=-\nabla
n(x,t)h(x,t)-\nonumber\\
\left[  \nabla V(x)-f(t)\right]  J(x,t)-\gamma k_{B}n(x,t)\left[
T(x,t)-T_{o}\right]  \label{tempera_balance}%
\end{gather}
where we have defined the heat flow as%
\begin{equation}
h(x,t)=\frac{1}{2}m\int u^{3}\phi_{x}(u,t)du.\label{heat_flow}%
\end{equation}
For times $t\gg\gamma^{-1}$, Eq. (\ref{heat_flow}) reduces to
\begin{gather}
k_{B}n(x,t)\left[  T(x,t)-T_{o}\right]  =\ -\gamma^{-1}\nabla
n(x,t)h(x,t)-\nonumber\\
\gamma^{-1}\left[  \nabla V(x)-f(t)\right]  J(x,t).\label{temp_qe}%
\end{gather}
In the particular case of local equilibrium $J(x,t)=h(x,t)=0$ which would lead
to $T(x,t)=T_{o}$. If, on the other hand, we subsitute the expression of
$J(x,t)$ given through eq. (\ref{current3}) into Eq. (\ref{temp_qe}) and using
Eq. (\ref{temp_qe}), up to order $\gamma^{-2}$ we obtain
\begin{gather}
k_{B}n(x,t)\left[  T(x,t)-T_{o}\right]  =\ -\gamma^{-1}\nabla
n(x,t)h(x,t)+\nonumber\\
\gamma^{-2}\left\{  mn(x,t)\left[  \nabla V(x)-f(t)\right]  ^{2}-k_{B}%
T_{o}\left[  \nabla V(x)-f(t)\right]  \nabla n(x,t)\right\}  ,\label{est_temp}%
\end{gather}
which after performing an average over $x$-space leads to\textbf{ }
\begin{equation}
k_{B}T(t)\simeq k_{B}T_{o}+m\gamma^{-2}\left\langle \nabla
V(x)-f(t)\right\rangle ^{2}.\label{temperature2}%
\end{equation}
Here $T(t)=\left\langle T(x,t)\right\rangle $ and we have neglected a term
$k_{B}T_{o}\left\langle \nabla\left[  \nabla V(x)-f(t)\right]  \right\rangle
$. Relation (\ref{temperature2}) means that for sufficiently large gradients
and forces applied on the system, its temperature will in general differ from
that of the heat bath. On the contrary, it must be emphasized that for small
gradients and forces the quasi-equilibrium temperature $T(t)$ given by Eq.
(\ref{temperature2}) reduces to the bath temperature $T_{o}$ implying that the
system reaches local equilibrium. This is precisely the hypothesis of small
forces and gradients which is usually behind the local equilibrium hypothesis
\cite{prigogine}. A correction to the temperature similar to
(\ref{temperature2}) has been previously obtained in a different context in
Ref. \cite{landau1}.

\section{Fluctuation theorems for local equilibrium and quasiequilibrium}

The transition from the intial state $x_{o}$ to the final state $x_{t}$, can
be modeled by a set of N coupled unimolecular chemical reactions having $x$ as
their reaction coordinate. This approach is based on the fact that the kinetic
or rate equation corresponding to a unimolecular chemical reaction represents
a gain and loss process which can be interpreted probabilistically as the
result of the balance of two opposite probability currents.

To compute the ratio between the forward and reverse probabilities of a path
between $x_{o}$ and $x_{t}$, we define a partition $t_{1},t_{2},.....,t_{N+1}$
of the entire time interval $\left[  0,t\right]  $, with $t_{1}=0$ and
$t_{N+1}=t$, which divides the process in N steps. The initial state for these
reactions coincides with $x_{o}$ and the final state with $x_{t}$, while the
intermediate states correspond with $x_{t_{2}},x_{t_{3}},...$. Thus, given the
probability $n_{i}$ of being in the state $x_{t_{i}}$ at time $t_{i}$, the
elementary kinetics is given by the set of equations
\begin{equation}
\frac{d}{d\tau}n_{i}=J^{i}-J^{i-1};\text{ \ }i=1,.....,N+1, \label{rate-eq}%
\end{equation}
where
\begin{equation}
J^{i}=v_{i+1,i}^{R}n_{i+1}-v_{i,i+1}^{F}n_{i};\text{ \ }J^{o}=J^{N+1}=0.
\label{current5}%
\end{equation}
Here, the symbols $F$ and $R$ stand for the forward and reverse transitions.
On the other hand, it must be stressed that the values of the rate constants
$v_{ij}^{F,R}(\tau)$ depend on whether the system is in a local equilibrium or
a quasi-equilibrium state. Therefore, since the reaction constant $r_{i,j}$
characterizing the asymmetry of the reversible reaction is defined by the
ratio between the rate constants, this will depend also on the state of the
system. According to its definition, $r_{i,j}$ gives us the ratio of the
probabilities of the forward and backward reactions.

In the quasi-stationary state $J^{1}=J^{2}=......=J^{N}=0$ and from Eqs.
(\ref{rate-eq}) and (\ref{current5}) we find
\begin{equation}
\lim\limits_{\tau\longrightarrow\infty}\frac{n_{i+1}(\tau)}{n_{i}(\tau)}%
=\frac{n_{i+1}^{o}(f)}{n_{i}^{o}(f)}=\frac{v_{i,i+1}^{F}(f)}{v_{i+1,i}^{R}%
(f)}=r_{i,i+1}(f) \label{rate-const}%
\end{equation}
where the upper index $o$ stands for the long time value of $n_{k}$, and
$r_{i,i+1}(f)$ is the partial equilibrium constant corresponding to the
\textit{i-th} reaction (step), which depends on time through $f$. Thus, for
the forward and reverse driven processes between $t_{1}$ and $t_{N+1}$, we
obtain the general relations
\begin{gather}
r_{1,N+1}(f)=r_{1,2}(f)r_{2,3}(f)...r_{N,N+1}(f)=\nonumber\\
\frac{n_{2}^{o}(f)}{n_{1}^{o}(f)}\frac{n_{3}^{o}(f)}{n_{2}^{o}(f)}%
.....\frac{n_{N+1}^{o}(f)}{n_{N}^{o}(f)}=\Pi_{i=1}^{N}\frac{n_{i+1}^{o}%
(f)}{n_{i}^{o}(f)}, \label{eff-rate-cons}%
\end{gather}
where now $n_{i}^{o}$ will have different dependencies on the temperature
depending on whether the system is in a local equilibrium or a
quasi-equilibrium state.

i) \textit{Local equilibrium}. By making use repeatedly of Eq. (\ref{eq-sol})
in Eq. (\ref{eff-rate-cons}) one obtains that the ratio between the forward
$prob^{F}(path)$ and reverse $prob^{R}(path)$ path probabilities is given by
\begin{gather}
\frac{prob^{F}(path)}{prob^{R}(path)}=r_{1,N+1}(f)=\exp\left[  \frac{m}%
{k_{B}T_{o}}\int_{0}^{t}\left[  -\nabla V(y)+f(\tau)\right]  \dot{y}%
\,d\tau\right]  =\nonumber\\
\exp\left[  -\frac{\Delta F}{k_{B}T_{o}}\right]  \exp\left[  \frac{W_{f}%
}{k_{B}T_{o}}\right]  \label{prob-rat}%
\end{gather}
where $W_{f}=m\int_{0}^{t}f(\tau)\dot{y}d\tau$ is the work due to the
nonconservative force in the forward path. Eq. (\ref{prob-rat}) constitutes
the ratio between the forward and backward path probabilities providing us the
Fluctuation-Theorem at local equilibrium \cite{crooks}. Additionally, by using
Eqs. (\ref{heat}) and (\ref{ext-entr}), we can rewrite Eq. (\ref{prob-rat}) in
the form
\begin{equation}
\frac{prob^{F}(path)}{prob^{R}(path)}=e^{-{\Delta F}/{k_{B}T_{o}}}%
e^{{\tilde{\sigma}}/{k_{B}}}, \label{prob-rat01}%
\end{equation}
where $\tilde{\sigma}=\int_{0}^{\tau}\frac{d_{ext}S}{dt}dt$ is the total
entropy produced by the non-conservative force along the process. Eq.
(\ref{prob-rat01}) expresses that the ratio of forward and backward
probabilities is related to the entropy production \cite{evans3}. Nonetheless,
it should be pointed out that $\tilde{\sigma}$ is the entropy change of the
system with its surroundings and therefore is of undefined sign. Another
consequence of Eq. (\ref{prob-rat}) is the following:
\begin{gather}
\left\langle \exp\left[  -\frac{W_{f}}{k_{B}T_{o}}\right]  \right\rangle =%
{\displaystyle\sum\limits_{path}}
prob^{F}(path)\exp\left[  -\frac{W_{f}}{k_{B}T_{o}}\right] \nonumber\\
=%
{\displaystyle\sum\limits_{path}}
prob^{R}(path)\exp\left[  \frac{-\Delta F}{k_{B}T_{o}}\right] \nonumber\\
=\exp\left[  \frac{-\Delta F}{k_{B}T_{o}}\right]  \label{non_eq-wor}%
\end{gather}
which constitutes the so-called nonequilibrium work relation \cite{jarzynski}.
Despite of the common belief, Eq. (\ref{non_eq-wor}) is only valid at local
equilibrium as was stated previously \cite{agusti2,mendeli}.

It is convenient to emphasize that the Fluctuation Theorem adopts so simple
and elegant expressions like Eqs. (\ref{prob-rat}) and (\ref{prob-rat01})
because we have derived them in the conditions of local equilibrium, i.e.,
when the fluctuation-dissipation regime is satisfied. Otherwise more fuzzy
expression is obtained, as we will show next.

ii) \textit{Quasi-equilibrium}. Following a similar analysis as in the
previous case, from the quasi-equilibrium solution given in Eq.
(\ref{quasi-equi}) we find%
\begin{gather}
\frac{prob^{F}(path)}{prob^{R}(path)}=r_{1,N+1}(f)\nonumber\\
=\exp\left[  \int_{0}^{t}\frac{m}{k_{B}T(\tau)}\left[  -\nabla V(y)+f(\tau
)\right]  \dot{y}d\tau\right]  \label{prob-rat2}%
\end{gather}
giving us the ratio between the forward and backward paths probabilities at
quasi-equilibrium. Eq. (\ref{prob-rat2}) constitutes a generalization of the
Fluctuation Theorem to the case in which the system not yet relaxed to local
equilibrium, but it is still at a quasi-equilibrium. As we have mentioned
previously, in far from equilibrium conditions, it is not generally correct to
assume that temperature of the system is that of the heat bath, and therefore
in far from equilibrium conditions Eq. (\ref{prob-rat2}) must be used in place
of (\ref{prob-rat}). From Eq. (\ref{prob-rat2}) it is not possible to obtain a
relation similar to (\ref{non_eq-wor}) because of the dependence of the
quasi-equilibrium temperature in $f$ and time along the path.

\section{Conclusions}

In this paper based on the thermodynamic theory of Brownian motion
\cite{mazur}, \cite{agusti3}, we have derived the Fokker-Planck equation for a
system of Brownian particles subjected to both conservative and
nonconservative forces. In this scenario we have obtained expressions for the
FT in local equilibrium and in quasi-equilibrium. Local equilibrium coincides
with the fluctuation-dissipation regime where the fluctuation-dissipation
theorem is satisfied and the intensive parameters characterizing the state of
the system coincide with those of the bath. In this case our expressions for
the FT agree with previous results in the literature.

Nonetheless, when arbitrary forces remove the system from equilibrium with the
bath and keep it out of equilibrium modifying the intensive parameters
characterizing the state of the system, the fluctuation-dissipation theorem is
no longer valid. However, if these external forces vary quite slowly in time,
one can define a state of quasi-equilibrium. Fluctuations around this
quasi-equilibrium state are characterized by a generalization of the FT in
which the existence of quasi-equilibrium state is taken into account. The
theory we have presented here enables us to derive steady state fluctuation
theorems for arbitrary forcing, something which we will do in a latter work.

\end{document}